# Quantum Communication and Computing With Atomic Ensembles Using Light-Shift Imbalance Induced Blockade

*M.S. Shahriar, G.S. Pati, and K. Salit*
*EECS Department, Northwestern University, Evanston, IL*

## Abstract

Recently, we have shown that for conditions under which the so-called light-shift imbalance induced blockade (LSIIB) occurs, the collective excitation of an ensemble of a multi-level atom can be treated as a closed two level system. In this paper, we describe how such a system can be used as a quantum bit (qubit) for quantum communication and quantum computing. Specifically, we show how to realize a C-NOT gate using the collective qubit and an easily accessible ring cavity, via an extension of the so-called Pellizzari scheme. We also describe how multiple, small-scale quantum computers realized using these qubits can be linked effectively for implementing a quantum internet. We describe the details of the energy levels and transitions in $^{87}$Rb atom that could be used for implementing these schemes.

When an atomic ensemble is excited, by a laser beam matched to a two-level transition (or a Raman transition) for example, it leads to a cascade of many states as more and more photons are absorbed[1,2,3]. In order to make use of an ensemble as a quantum bit (qubit), it is necessary to disrupt this cascade, and restrict the excitation to the absorption (and emission) of a single photon only. In principle, this can be achieved through the use of the so-called dipole blockade[4,5]. In order to make use of this blockade mechanism in a manner that is consistent with a quantum computing architecture, it is necessary to control the distribution of inter-atomic distances between each pair of atoms in the ensemble in a precise manner  Furthermore, in order to achieve long decoherence times, it is necessary to make use of dipole-blockade based on spin-spin coupling, which is necessarily much weaker than the optical dipole-dipole coupling. Recently, we have shown that a new type of blockade mechanism, based on the light-shift imbalance in a Raman transition, can overcome these constraints. The resulting system does not impose any constraint on the distribution of inter-atomic distance within an ensemble. Furthermore, no dipole-dipole coupling is necessary, so that a relatively low density system can be used.

In reference[6], we have shown how this light shift imbalance induced blockade (LSIIB) process enables one to treat the ensemble as a two level atom that undergoes fully deterministic Rabi oscillations between two collective quantum states, while suppressing excitations of higher order collective states. In this paper, we show how this transition can be used to realize a qubit embodied by the ensemble. Using multiple energy levels inside each atom, we show how the LSIIB enables the transfer of quantum information between neighboring ensembles, as well as the realization of a C-NOT gate. In effect, this represents a generalization of the so-called Pellizzari scheme for quantum information processing[7]. Furthermore, we show that the LSIIB can be used to link two separate quantum computers (QC), by transferring the quantum state of any ensemble qubit in one QC to any ensemble qubit in another QC. We discuss practical ways to implement this scheme using specific energy levels and transitions in $^{87}$Rb atoms, and propose experiments to demonstrate the feasibility of these protocols.

The significance of the light shift imbalance induced blockade (LSIIB) process can be summarized as follows: (a) It can be used to realize a deterministic quantum bit encoded in the collective-excitation states of an atomic ensemble. (b) Along with an easily accessible ring cavity, it can be used to realize a two-qubit gate (e.g., a C-NOT gate) between two ensemble-based qubits. (c) It can be used to transport, deterministically, the quantum state of an ensemble qubit from one location to another separated by macroscopic distances. (d) It can be used to establish a quantum-link between two ensembles-and-cavity based quantum computers. The scheme proposed here thus offers a robust technique for realizing a quantum internet without using the single-atom and ultra-short cavity based approaches[8,9,10].

Before proceeding, we summarize briefly the notations for describing the LSIIB process[6]. In the simplest version, we consider each atom in the ensemble to be a $\Lambda$-type three level system, with two metastable states ($|a\rangle$ and $|c\rangle$), and an optically excited state ($|e\rangle$), as shown in figure1. The collective states of interest are defined as follows[6]:

$$|A\rangle = |a_1, a_2, \cdots, a_N\rangle \,;\; |G_1\rangle = \frac{1}{\sqrt{N}} \sum_{j=1}^{N} |a_1, a_2, \cdots, g_j, \cdots a_N\rangle \,;\; |C_1\rangle = \frac{1}{\sqrt{N}} \sum_{j=1}^{N} |a_1, a_2, \cdots, c_j, \cdots a_N\rangle$$

$$|G_2\rangle = \frac{1}{\sqrt{{}^N C_2}} \sum_{j,k(j\neq k)}^{{}^N C_2} |a_1, a_2, \cdots, g_j, \cdots, g_k, \cdots a_N\rangle \,;\; |C_2\rangle = \frac{1}{\sqrt{{}^N C_2}} \sum_{j,k(j\neq k)}^{{}^N C_2} |a_1, a_2, \cdots, c_j, \cdots, c_k, \cdots a_N\rangle$$

$$|G_{1,1}\rangle = \frac{1}{\sqrt{2 {}^N C_2}} \sum_{j,k(j\neq k)}^{2 {}^N C_2} |a_1, a_2, \cdots, g_j, \cdots, c_k, \cdots a_N\rangle$$

$$|G_{2,1}\rangle = \frac{1}{\sqrt{Z}} \sum_{(j\neq k \neq l)}^{Z} |a_1, a_2, \cdots, g_j, \cdots, g_k, \cdots, c_l, \cdots a_N\rangle$$

$$|G_{1,2}\rangle = \frac{1}{\sqrt{Z}} \sum_{(j\neq k \neq l)}^{Z} |a_1, a_2, \cdots, g_j, \cdots, c_k, \cdots, c_l, \cdots a_N\rangle \quad [Z = 3 {}^N C_3]$$

(1)

The relevant coupling rates between these collective states are also illustrated in figure 2. Note that for large detunings, the excitations to the intermediate states $|G_1\rangle$ and $|G_{1,1}\rangle$ are small, so that higher order states such as $|G_2\rangle$ and $|G_{2,1}\rangle$ can be ignored. The remaining system looks very similar to the five-level system considered in section B above. (Parenthetically at this point, note that the coupling between $|G_1\rangle$ and $|C_1\rangle$ does not scale with $\sqrt{N}$, unlike the coupling between $|A\rangle$ and $|G_1\rangle$, which scales as $\sqrt{N}$.)

In the rotating wave transformation frame, the truncated, six level Hamiltonian, in the bases of $|A\rangle$, $|G_1\rangle$, $|C_1\rangle$, $|G_{1,1}\rangle$, $|C_2\rangle$ and $|G_{1,2}\rangle$ is given by (the justification for not including the state $|C_3\rangle$ will be made by showing that the excitation to $|C_2\rangle$ can be suppressed, thus in turn making the amplitude of $|C_3\rangle$ insignificant):

$$H = \begin{bmatrix} \Delta/2 & \sqrt{N}\Omega_1/2 & 0 & 0 & 0 & 0 \\ \sqrt{N}\Omega_1/2 & -\delta & \Omega_2/2 & 0 & 0 & 0 \\ 0 & \Omega_2/2 & -\Delta/2 & (\sqrt{N-1})\Omega_1/2 & 0 & 0 \\ 0 & 0 & (\sqrt{N-1})\Omega_1/2 & -(\delta+\Delta) & \sqrt{2}\Omega_2/2 & 0 \\ 0 & 0 & 0 & \sqrt{2}\Omega_2/2 & -3\Delta/2 & (\sqrt{N-2})\Omega_1/2 \\ 0 & 0 & 0 & 0 & (\sqrt{N-2})\Omega_1/2 & -(\delta+2\Delta) \end{bmatrix} \quad (2a)$$

where the detunings are defined as: $\delta \equiv (\delta_1 + \delta_2)/2$ and $\Delta \equiv (\delta_1 - \delta_2)$.

If the detunings are large compared to the transition rates, we can eliminate states $|G_1\rangle$, $|G_{1,1}\rangle$ and $|G_{1,2}\rangle$ adiabatically. Under this condition, the effective Hamiltonian for the three remaining states ($|A\rangle$, $|C_1\rangle$, and $|C_2\rangle$) are given by (assuming $\delta \gg \Delta$):

$$\widetilde{H} = \begin{bmatrix} \varepsilon_A + \Delta/2 & \Omega_{Ro}/2 & 0 \\ \Omega_{Ro}/2 & \varepsilon_{C1} - \Delta/2 & \sqrt{2(N-1)/N}\,\Omega_{Ro}/2 \\ 0 & \sqrt{2(N-1)/N}\,\Omega_{Ro}/2 & \varepsilon_{C2} - 3\Delta/2 \end{bmatrix} \quad (2b)$$

where $\varepsilon_A$, $\varepsilon_{C1}$, and $\varepsilon_{C2}$ are the light shifts of the states $|A\rangle$, $|C_1\rangle$, and $|C_2\rangle$, respectively. and $\Omega_{Ro} \equiv (\sqrt{N}\Omega_1\Omega_2)/(2\delta)$. To first order, these light shifts are given by $\varepsilon_A = N\Omega_1^2/4\delta$, $\varepsilon_{C_1} = [\Omega_2^2 + (N-1)\Omega_1^2]/4\delta$, and $\varepsilon_{C_2} = [2\Omega_2^2 + (N-2)\Omega_1^2]/4\delta$, and are balanced, in the sense that $(\varepsilon_{C1} - \varepsilon_A) = (\varepsilon_{C2} - \varepsilon_{C1})$. This means that if the explicit two-photon detuning, $\Delta$, is chosen to make the Raman transition between $|A\rangle$ and $|C_1\rangle$ resonant (i.e., $\Delta = (\varepsilon_{C1} - \varepsilon_A)$), then the Raman transition between $|C_1\rangle$ and $|C_2\rangle$ also becomes resonant. This balance is broken when the light shifts are calculated to second order, and the blockade shift is then given by

$$\Delta_B \equiv (\varepsilon_{C2} - \varepsilon_{C1}) - (\varepsilon_{C1} - \varepsilon_A) = -(\Omega_2^4 + \Omega_1^4)/(8\delta^3). \quad (3)$$

With the proper choice of two-photon detuning ($\Delta = (\varepsilon_{C1} - \varepsilon_A)$) to make the Raman transition between $|A\rangle$ and $|C_1\rangle$ resonant, the effective Hamiltonian (after shifting the zero of energy, and assuming $N \gg 1$) is now given by:

$$\widetilde{H} = \begin{bmatrix} 0 & \Omega_{Ro}/2 & 0 \\ \Omega_{Ro}/2 & 0 & \Omega_{Ro}/\sqrt{2} \\ 0 & \Omega_{Ro}/\sqrt{2} & \Delta_B \end{bmatrix} \quad (4)$$

This form of the Hamiltonian shows clearly that when $[\Omega_{Ro}/\sqrt{2}] \ll \Delta_B$, the coupling to the state $|C_2\rangle$ can be ignored. As such, the collective excitation process leads to a Rabi oscillation in an *effectively closed two level system* consisting of $|A\rangle$ and $|C_1\rangle$. This is the LSIIB in the context of ensemble excitation, and is the key result upon which most of the protocols we propose here are based.

Thus, we can now represent a quantum bit by this effectively closed two level system. In the process, we have also shown how to perform an arbitrary single qubit rotation, an essential pre-requisite for quantum computing. The essence of the LSIIB for ensemble excitation can be summarized as follows. Whenever we have a three level optically off-resonant transition for the individual atoms, this can be translated into a corresponding off-resonant three-level transition involving collective states, which in turn is reduced to an effective two-level transition, as illustrated in figure 3. In order for this to hold, the primary constraint is that, for the collective states, the Rabi frequency on one leg must be much bigger than the same for the other.

At this point, it is instructive to consider a specific numerical example. Choosing the natural decay rate, $\Gamma$, of the excited state to be unity, assume $\Omega_2=100$, $\Omega_1=\eta \ll 1$, and $\delta=1000$. This corresponds to a value of $\Delta=\delta_1-\delta_2=\varepsilon_{C1}-\varepsilon_A \cong 2.5$, satisfying the condition that $\delta \gg \Delta$. The value of the blockade shift is $\Delta_B \cong -1/80$. To be concrete, we satisfy the requirement that $\Omega_{Ro}/\sqrt{2} \ll \Delta_B$ by demanding that $(\Omega_{Ro}/\sqrt{2})=\Delta_B/10$. This translates to the condition that $\eta\sqrt{N}=0.035$. Note that these parameters satisfy the constraint that $\delta \gg \Omega_2$, and $\delta \gg \Omega_1\sqrt{N}$. The acceptable range for N will be dictated by the choice of $\eta$, or vice-versa, depending on the particular experiment at hand. As an example, for $\eta=10^{-3}$, we need $N \cong 1200$. To see whether such a range of parameters are potentially realistic, let us consider an ensemble of cold $^{87}$Rb atoms caught in a FORT trap, excited by control beams with a cross-sectional diameter of about 200 µm. Assuming that the transitions employed dipole matrix element amplitudes that are half as strong as those of the strongest transitions, the power needed for the $\Omega_2$ beam is about 100 mW, and that for the $\Omega_1$ beam is about 10 pW. The time for a π-transition going from $|A>$ to $|C_1>$ is about 50 µsec. Given that the decoherence time in a FORT can be of the order of minutes, as many as $10^6$ qubit operations can be carried out at this rate. The number of photons in the $\Omega_1$ is close to 2000, so that its treatment as a classical beam is valid.

The LSIIB process can be used to perform two qubit operations between two ensembles. In particular, we show here how a C-NOT gate can be realized in such a system. The process we describe is essentially the ensemble version of the well-known Pellizzari scheme[7]. Figure 4(top) shows the basic configuration involving only two ensembles and a cavity. A more general scheme that can couple the nearest neighbor ensembles in a large array of ensembles is described schematically later on. In what follows, the two ensembles will be denoted as Ensembles I and II, respectively. Figure 4(bot) shows the necessary energy levels in each of the atoms. The transition corresponding to the cavity mode is also illustrated, with a vacuum Rabi frequency of $g_c$. Later on, we will discuss how an alkali atom such as $^{87}$Rb may be used to implement these energy levels.

We assume that both ensembles are prepared in the $|A>$ state. For the other states, the only ones that will be relevant are the ones that are of the following form:

$$|Q_1\rangle \equiv \frac{1}{\sqrt{N}}\sum_{j=1}^{N}|a_1,a_2,\cdots,q_j,\cdots a_N\rangle; \quad q=s,b,c,d,e,f,g \text{ or } h \text{ (excluding } a\text{)} \quad (5)$$

We now describe the C-NOT gate operation in multiple steps.

***Step 1: Initialize each ensemble in an arbitrary quantum superposition***: At t=0, we assume that each ensemble is in state $|A>$: $|\psi\rangle_I = |A\rangle_I$; $|\psi\rangle_{II} = |A\rangle_{II}$. This can be achieved, in principle, by optical pumping in a real system. *Operations for Ensemble 1:* The transitions to be used for performing the single qubit rotation is shown in figure 5. With $\Omega_2 \ll \Omega_1$, we apply these two pulses simultaneously for duration $T_1$. The state of ensemble I (E-I) is then given by: $|\psi\rangle_I = Cos[\Omega_{Ro}T_1/2]|A>_I + iSin[\Omega_{Ro}T_1/2]|C_1>_I \equiv \alpha|A>_I + \beta|C_1>_I$. Here, $\Omega_{Ro}$ is as defined in eqn. 2 above. This achieves the goal of producing an arbitrary quantum state in E-I. *Operations for Ensemble II:* For the second ensemble, we apply the same transitions as in figure 5a, but for a duration $T_2$ such that $\Omega_{Ro}T_2 = \pi$. This corresponds to a pi-pulse, and Ensemble II (E-II) is now in state $|C_1>$. This is now followed by the excitation pulses shown in figure 5b. Note in particular that the coupling rates for the collective states no longer includes the √N factor. The pulses at $\Omega_1^/$ and $\Omega_2^/$ are applied for a duration $T_3$ (satisfying the condition that $\Omega_2^/ \ll \Omega_1^/$), so that the state for E-II is now given by: $|\psi\rangle_{II} = Cos[\Omega^/T_3/2]|C>_{II} + iSin[\Omega^/T_1/2]|D>_{II} \equiv \xi|C>_{II} + \eta|D>_{II}$. Thus, we now have prepared E-II in an arbitrary qubit state. Finally, note that the cavity so far is in the ground state with no photons: $|\psi\rangle_C = |0\rangle_C$.

***Step 2: Transfer State of E-I to the cavity***: The transitions to be employed for this step is illustrated in figure 6a. The pulse at $\Omega_I$ is applied for a duration $T_4$ such that $(\sqrt{N}\Omega_I g_C/2\delta)\cdot T_4 = \pi$. As a result of this pi-pulse, the states of E-II remains unchanged, but the states of E-I and the cavity become as follows: $|\psi\rangle_C = \alpha|0\rangle_C + \beta|1>_C$; $|\psi\rangle_I = |C_1\rangle_I$. This accomplishes the goal of transferring the quantum state of E-I to the cavity.

***Step 3: Transfer State of the cavity to E-II***: The transitions used for this step, in E-II, is illustrated in figure 6b. Note again that the transition rates for coupling the collective states now do not include the √N factor.

The pulse at $\Omega_{II}$ is applied for a duration $T_5$ such that $(\Omega_{II} g_C / 2\delta) \cdot T_5 = \pi$. As a result of this pi-pulse, the states of E-I remains unchanged, but the states of E-II and the cavity become as follows: $|\psi\rangle_C = |0\rangle_C$; $|\psi\rangle_{II} = \alpha\xi|S_1\rangle_{II} + \xi\beta|C_1\rangle_{II} + \eta\alpha|B_1\rangle_{II} + \eta\beta|D_1\rangle_{II}$. This accomplishes the goal of transferring the quantum information from the cavity to E-II. Of course, steps 2 and 3 together achieves the objective of transferring all the quantum information from both ensembles into ensemble II only.

*Step 4: Perform "Effective C-NOT transition" inside E-II:* Using the transitions shown in figure 5b, we apply a pi-pulse to E-II, which exchanges the amplitudes of $|C_1\rangle_{II}$ and $|D_1\rangle_{II}$. As a result, while E-I and the cavity stays unchanged, the state of E-II now becomes: $|\psi\rangle_{II} = \alpha\xi|S_1\rangle_{II} + \xi\beta|D_1\rangle_{II} + \eta\alpha|B_1\rangle_{II} + \eta\beta|C_1\rangle_{II}$.

*Step 5: Transfer E-II Quantum State Partly to the Cavity:* Using the transitions shown in figure 6b, we now apply another pi-pulse in E-II. As a result, while E-I stays unchanged, the cavity and the E-II now become entangled: $|\psi\rangle_{II-C} = \alpha\xi|C_1\rangle_{II}|1\rangle_C + \xi\beta|D_1\rangle_{II}|0\rangle_C + \eta\alpha|D_1\rangle_{II}|1\rangle_C + \eta\beta|C_1\rangle_{II}|0\rangle_C$.

*Step 6: Transfer Cavity State back to E-I:* Using the transitions shown in 3a, we now apply another pi-pulse in E-I. As a result, the cavity returns to the zero-photon state, and E-I and E-II are now entangled: $|\psi\rangle_C = |0\rangle_C$; $|\psi\rangle_{I-II} = \alpha\xi|A\rangle_I|C_1\rangle_{II} + \eta\alpha|A\rangle_I|D_1\rangle_{II} + \beta\xi|C_1\rangle_I|D_1\rangle_{II} + \beta\eta|C_1\rangle_I|C_1\rangle_{II}$. This is to be compared with the direct-product state at the beginning of the protocol: $|\psi\rangle_{I-II}(init) = |\psi\rangle_I|\psi\rangle_{II} = \alpha\xi|A\rangle_I|C_1\rangle_{II} + \alpha\eta|A\rangle_I|D_1\rangle_{II} + \beta\xi|C_1\rangle_I|C_1\rangle_{II} + \beta\eta|C_1\rangle_I|D_1\rangle_{II}$. Thus, the quantum state of E-II has been flipped only for the case where the quantum state of E-I is in $|C_1\rangle$, which, of course, corresponds to the C-NOT operation.

The whole operation for realizing inter-processor communication is illustrated in figure 7. Briefly, each QC is modeled as an array of ensembles that can be moved linearly, using several different possibilities, including quantum motors or sliding standing waves through a cavity[11,12,13,14,15]. It is also possible to consider an implementation where race-track shaped microcavities are used on alternating sides, with the atom seeing the cavity mode through evanescent modes. Note also that while we are concentrating primarily on ensembles of neutral atoms here, there are solid material and systems that display the kind of narrow Raman transitions necessary for implementing such a scheme[16,17,18], although we have not yet worked out the details for an explicit solid state system.

The objective of the communication link is to transfer the quantum state of the ensemble (Q1) on the edge of one QC (QC1) to the ensemble (Q2) on the edge of the other QC (QC2). To start with, assume that Q1 is in an arbitrary quantum state (the protocol would work just as well in the more useful case when this atom is entangled with the rest of QC1), and Q2 is in its ground state: $|\psi\rangle_{Q1} \equiv \alpha|A\rangle_{Q1} + \beta|C_1\rangle_{Q1}$; $|\psi\rangle_{Q2} = |A\rangle_{Q2}$. The transitions to be used for the transfer process are illustrated in the bottom of figure 7. Note that here we are using the same set of transition as in figure 5a, except that instead of using a semi-classical mode to couple the |g> to |c> transition, we use a free-space single photon mode. The parameters of this mode (such as the vacuum Rabi frequency $g_F$) is determined essentially by the pump pulse[3,18]. We now apply a pulse for a duration $\tau_R$ such that: $(\sqrt{N}\Omega_{1C} g_F / 2\delta) \cdot \tau_R = \pi$. The result would be a transfer of the quantum state of Q1 to the free-space photon mode (in essentially the same way as in step-2 of the C-NOT protocol). A fiber delay line will allow one to let this photon state leave QC1 completely. The pump pulse on the Q2 will be timed with the anticipated arrival of this photon state (this can be done, since the delay between the photon state and the pump in Q1 can be pre-calibrated as well as calculated). This pump will then be applied for a duration $\tau_W$ such that: $(\sqrt{N}\Omega_{2C} g_F / 2\delta) \cdot \tau_W = \pi$. Note that the presence of the $\sqrt{N}$ factor makes both of these processes fast enough even for a free-space photon mode. After the second pulse, the free-space photon mode will be in the zero photon state, and the quantum states of Q1 and Q2 would be: $|\psi\rangle_{Q2} \equiv \alpha|A\rangle_{Q2} + \beta|C_1\rangle_{Q2}$; $|\psi\rangle_{Q1} = |A\rangle_{Q1}$. It is obvious that if Q1 were entangled with the rest of QC1, then Q2 will now be entangled with the rest of QC1. This is, of course, all that is needed to link the two quantum computers.

In order to see the feasibility of these operations, it is important to consider the relevant cavity parameters. When the ensemble qubit is interacting with the cavity mode (e.g., in Step 2 above), the vacuum coupling rate is enhanced by a factor of $\sqrt{N}$, so that the cavity parameters for reaching the strong-coupling regime is significantly relaxed. To illustrate, consider the situation where a cavity holds an ensemble rubidium qubit of $N$ atoms, with a length of $L$, and an effective mode diameter of $d$ at the center. The single-photon electric field amplitude is then given by $\sqrt{(2\hbar\omega)/(\varepsilon_o V)}$ where $V \simeq (\pi/4)d^2 L$ is the effective mode volume[19], and ω is the frequency of the photon. For the rubidium atom, with $N$=1, the corresponding vacuum rabi frequency is proportional to this field amplitude, is given by $g_o \simeq 54.25\Gamma$ (where $\Gamma \simeq 2\pi \cdot 6 \cdot 10^6$ sec$^{-1}$ is the natural linewidth) for $L_o$=40μm and $d_o$=5μm[20]. If we assume a matched pair of essentially lossless mirrors, each with an intensity transmittivity of 1.2X10$^{-6}$, the corresponding cavity finesse is about 2.6*10$^6$, with a free spectral range (FSR) of about 3.8*10$^{12}$ Hz[21]. This yields a cavity decay rate, $\gamma_o$ (HWHM), of about 0.12Γ, so that the cavity lifetime, $\tau_o \equiv \gamma_o^{-1}$, is about 222 nsec. From the discussion above, it is easy to see that these parameters scale as follows: $g = g_o (d_o/d)(L_o/L)^{1/2} N^{1/2}$, and $\tau = \tau_o (T_o/T)(L/L_o)$.

Consider next a situation where the number of atoms in the collective qubit is 3000. Using the scaling laws stated above, we see that the vacuum rabi frequency will remain unchanged if the length is increased by a factor of 3000 (L=12 cm). Here, we consider a cavity length of L=5 cm, so that the vacuum rabi frequency is $g_o \simeq 84.04\Gamma$, and the FSR is 3.0*10$^9$ Hz. For the same finesse as above, the cavity decay rate is now only 9.55X10$^{-5}$ Γ, with the corresponding cavity lifetime being close to 0.3 msec. With expected improvements in the mirror qualities, the cavity storage time may possibly become even longer. If we choose the strength of $\Omega_1$ to be 7X10$^{-4}$Γ, with a common-mode detuning of δ=10$^3$Γ, the constraints for the LSIIB are easily satisfied. The time needed for a π-pulse is about 50 μsec, which is considerably shorter than the cavity lifetime. In this context, we point out that a variant of the Pellizari scheme[7] --- which makes use of the so-called Cavity Dark State[18] --- can also be used to implement a CNOT gate. When this approach is used, the cavity remains virtually unpopulated during the operation, thereby reducing the constraint on the cavity lifetime. Another important consequence of such a long cavity is that it allows one to use a ring-configuration rather than a conventional Fabry-Perot configuration, thereby avoiding the formation of standing waves in the cavity. The long cavity also makes it potentially easier to access the ensemble with the external control beams.

The energy levels and selection rules necessary for implementing these schemes can be accommodated by using, for example, Zeeman sub-levels in the D1 transition manifold in $^{87}$Rb atoms. One explicit construct is shown in figure 8. A non-zero magnetic field is used to lift the degeneracy between the state |a> and states {|s> and |b>}. This is necessary in order to ensure, for example, that during the process outlined in figure 5a, there is no Raman coupling between |s> and |c> (or between |b> and |d>). Similar constraints also apply for processes illustrated in figure 6a. For the process in figure 6b, the opposite constraint applies (that is |s> to |c> as well as |b> to |d> are allowed, but |a> is not coupled due to Raman detuning). The amount of magnetic field to be applied can be rather small, (typically less than a Gauss), with the constraint that the Zeeman shift should be larger than the Raman transition linewidths. The second order Zeeman shift for the |a> state would be very small in this case, and can be taken into account in choosing the relevant Raman resonance conditions.

Note that the g-factors for the F=1 and F=2 levels are equal and opposite, which ensures that the s-c transition and the b-d transitions can be simultaneously Raman resonant. Note also that the g-factors for the F'=1 and F'=2 are also equal and opposite of each other, and are a factor of a three smaller than those for the F=1 and F=2 levels. More importantly, because the Raman transitions are far detuned from the F' sublevels optically, the Zeeman shifts of the sublevels for F'=1 and F'=2 are of virtually no significance.

The energy difference between the F'=1 and F'=2 is more than 816 MHz. Thus, for processes that are detuned below resonance with respect to the F'=1 (e.g., for the process in figure 6b) sublevel will not be influenced significantly by the F'=2 level. By the same token, for processes that are detuned above resonance with respect to the F'=2 (e.g., for the process in figure 5a) sublevel will not be influenced significantly by the F'=1 level. This consideration was taken into account in designing the protocols, so that for some operations the detuning is negative (e.g., fig. 2a), while for others the detuning is positive (e.g., fig. 3b).

Finally, note that in order to allow for the selection rules assumed in the protocols, it is necessary to use a combination of σ$_+$, σ$_-$, and π polarized light. This can be easily accommodated, since two orthogonal directions are free for applying classical field even in the presence of a cavity.

Using a configuration where a cluster of atoms are trapped using a FORT beam it should be possible to demonstrate the ensemble-based LSIIB process, without using a cavity. Note that, for an ensemble, this process occurs in a pure Λ system. This can be realized by using the states |1>, |2> and |3> of figure 8b. The atoms will first

be optically pumped into state |1>. The two-opposite circularly polarized beams, detuned from the F'=1 level, will then be used to excite the Raman transition between |1>, |2> and |3>. Since the off-resonant excitation produces negligible population in |2>, no repump beam would be necessary.

The evidence of LSIIB in this case, of course, is not as simple as in the case of single atoms[6]. What is necessary in this case is to demonstrate that one and only one atom (on average) can be excited to level |3>. In order to test this, one can operate the Raman transition process in a pulsed mode, with a variable pulse duration. After applying a Raman pulse, the beams exciting the 1-2-3 transition would be turned off, and another Raman transition would be applied to transfer the atom(s) in state |3> to the F=2, $m_F$=2 level. A cycling transition that couples this state to the F=3, $m_F$=3 level in the D2 manifold would then be used to collect fluorescence produced from this atom. The fluorescence could be collected by a high numerical aperture imaging system, and then transported through a fiber to a single photon counter. By taking into account all the experimental parameters, one should thus be able to determine the number of atoms that were excited to state |3>. The proof of ensemble-LSIIB would be established by showing that this never exceeds 1 for any pump-pulse duration.

The specific geometry of the experimental setup that could be used to demonstrate the ensemble-LSIIB based CNOT gate is illustrated schematically in figure 9a. The FORT beams shown could be produced from a single beam with imaging optics. Using dichroic mirrors, additional beams (to be used for producing the control pulses for each ensemble) could be brought in, parallel to the FORT beams. Additional control beams would be brought in a direction perpendicular to both the cavity axis and the FORT axis. This degree of freedom would enable us to satisfy the polarization selection rules.

In order to keep the process as clean as possible, it is advisable to consider first the simplest non-trivial version of the C-NOT gate. Recall the most general form of the input (unentangled) state of the two ensembles, and the corresponding entangled state after the C-NOT operation, one could choose the parameters such that α=β=1/√2, ξ=1 and η=0. The entangled state produced by the C-NOT gate would therefore be given by: $|\psi\rangle_{I-II} = [|A\rangle_I |C_1\rangle_{II} + |C_1\rangle_I |D_1\rangle_{II}]/\sqrt{2}$. Here, Ens-I is the ensemble trapped by FORT1, and Ens-II is the one trapped by FORT2, for example.

In order to demonstrate that such a state has indeed been produced, one can proceed in several different ways. One option is to apply the pulse described in figure 7, with the pump beam parallel to the direction of FORT1 (which holds Ens-I). If Ens-I is in state |A>, then application of a pulse so that $(\sqrt{N}\Omega_{1C} g_F / 2\delta) \cdot \tau_R = \pi$ will produce a single photon in the free-space mode (characterized by the single-photon Rabi-frequency of $g_F$). As described in detail previously, the field at $\Omega_{1C}$ is coupled to the |a>-|g> transition with a large positive detuning, and the field for $g_F$ is coupled to the |c>-|g> transition with the same detuning. The frequency of the free-space mode is selected automatically in order to satisfy the two-photon transition condition. If Ens-I is in state |$C_1$>, then application of the same pulse will not produce a photon in this free space mode. The photon in this mode could be detected by using an additional fiber-coupled single-photon detector, preceded by polarization and spectral filters.

In parallel to this process, a pump-beam (with Rabi-frequency $\Omega'_{2C}$) would be applied along FORT2 (which houses Ens-II), coupled to the |c>-|h> transition, with a large, positive detuning. The corresponding free-space mode (with vacuum Rabi frequency of $g'_F$), coupled to the |d>-|h> transition, would be automatically chosen to have the same detuning, in order to satisfy the two-photon resonance condition. This photon would be passed through the same set of three filters, and then detected by a fiber-coupled photon counter. As before, application of a pi-pulse such that $(\Omega'_{2C} g'_F / 2\delta) \cdot \tau'_R = \pi$ would produce a single-photon in the $g'_F$ only if Ens-II is in state |$C_1$>; otherwise, there would be no photons in this mode. Note, again, that there is no √N factor involved in this expression. A co-incidence between the photon counts found in the $g_F$ and the $g'_F$ modes will prove that the C-NOT gate was applied correctly, and confirm the entanglement between the ensembles. The transitions are illustrated schematically in figure 9b.

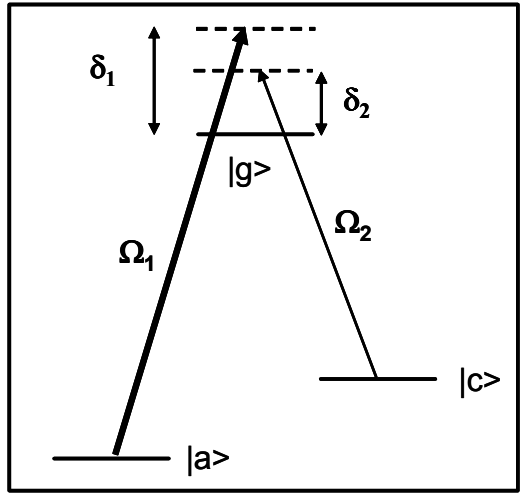

**Figure 1:** *Schematic illustration of a three level transition in each atom in an ensemble*

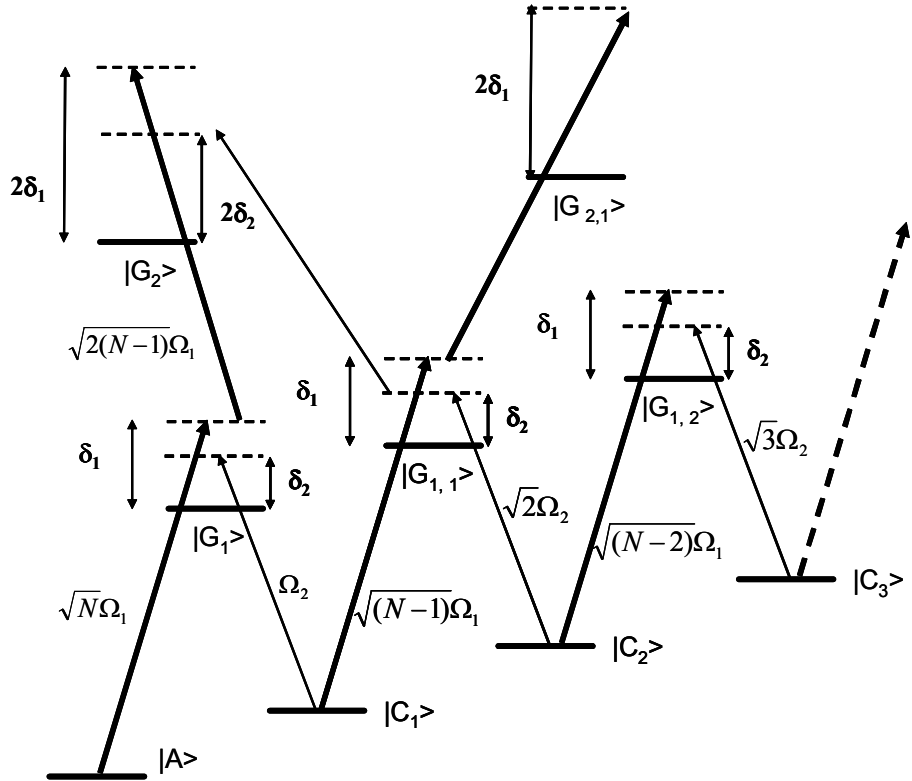

**Figure 2:** *Schematic illustration of the relevant collective states and the corresponding transition rates. See text for details*

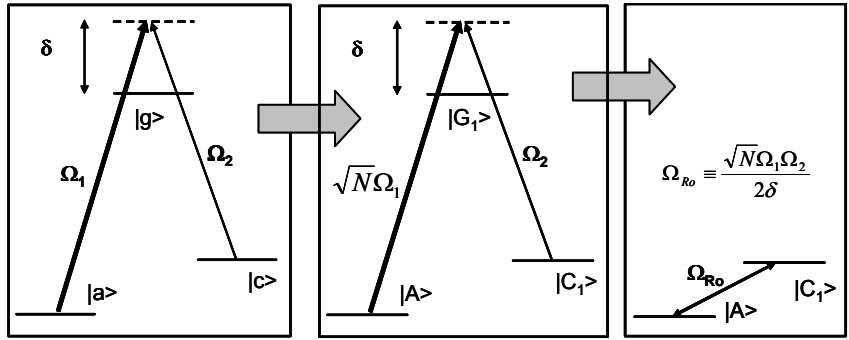

**Figure 3:** *Summary of the LSIIB process in an ensemble. See text for details.*

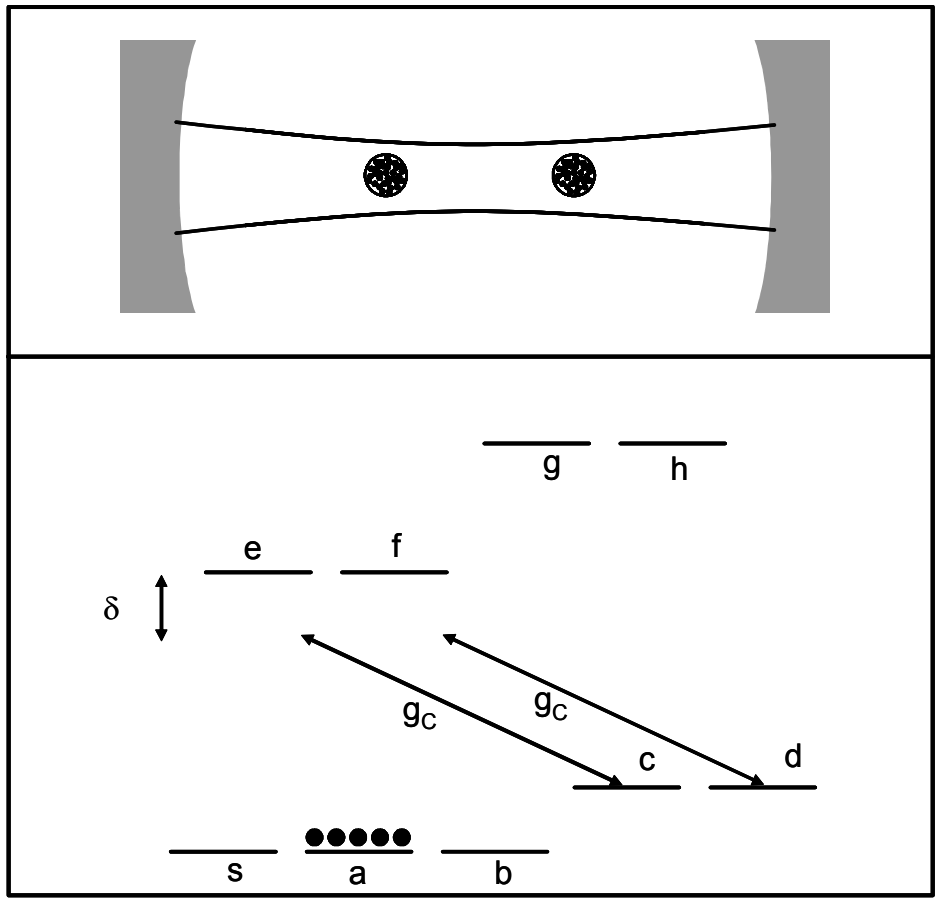

**Figure 4:** *Illustration of the basic configuration for coupling two ensembles (top), and the requisite energy levels for each atom (bot). See text for details.*

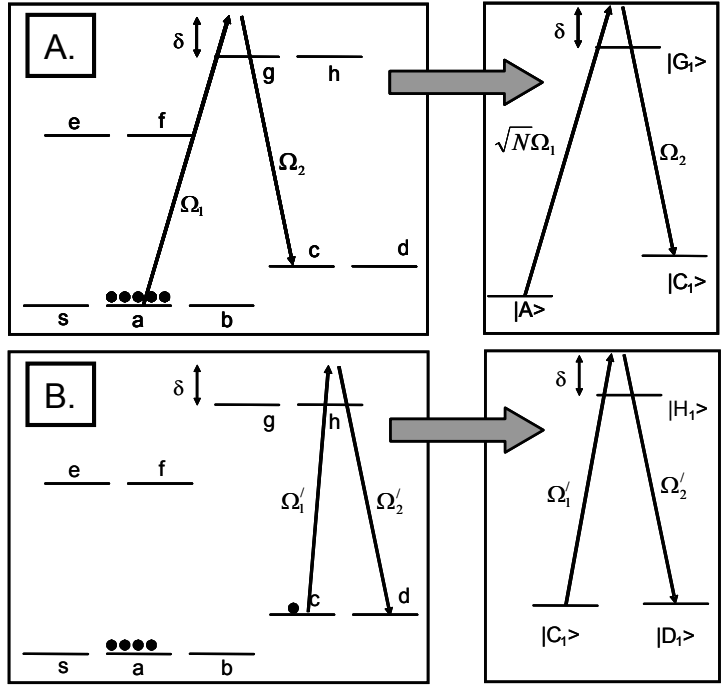

**Figure 5:** *(a) Schematic illustration of the single qubit state preparation in ensemble I. (b) Schematic illustration for the single-qubit operation for ensemble II. See text for details*

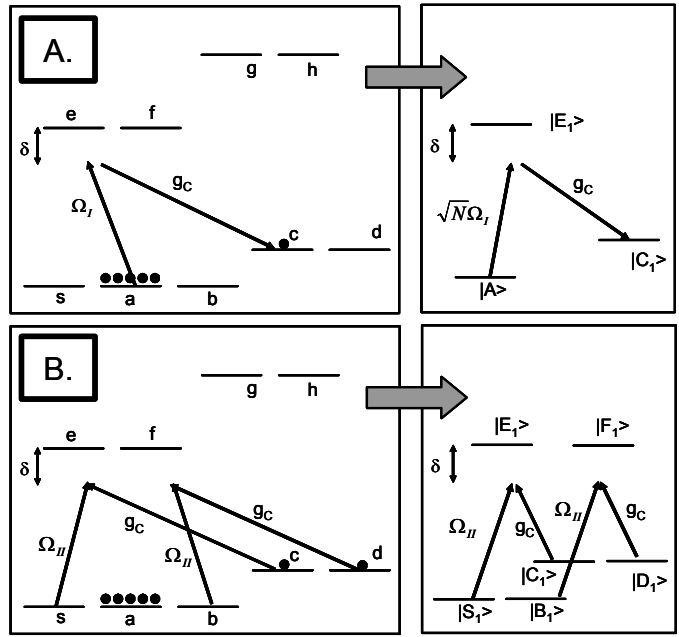

**Figure 6:** *(a) Illustration of the excitation step used to transfer the state of E-I to the cavity. (b) Schematic illustration of the pulses inside E-II to transform the cavity state to E-II. See text for details.*

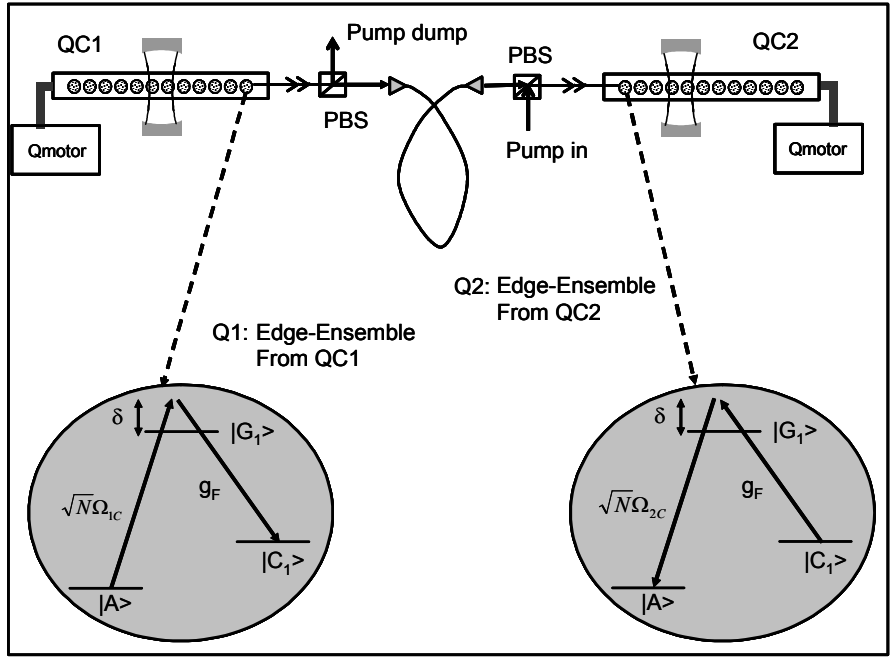

**Figure 7:** *Schematic illustration of quantum communication between two quantum computer. See text for details.*

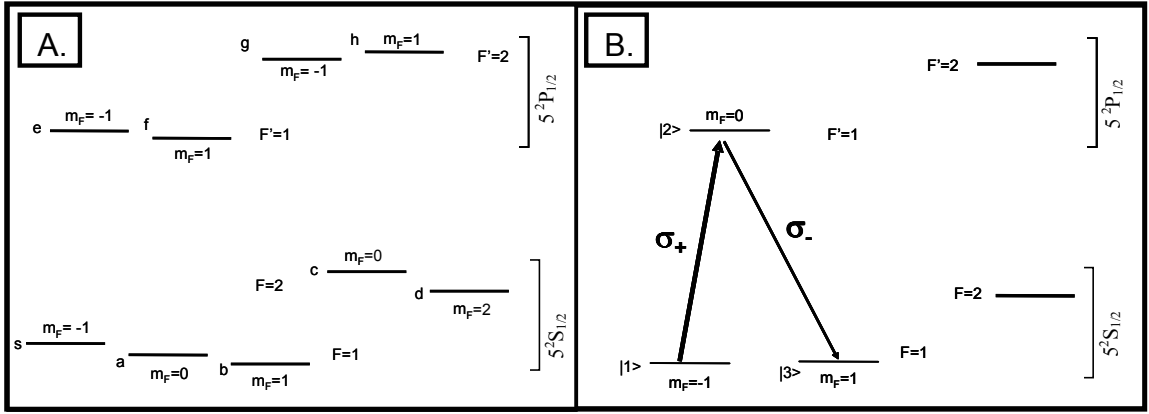

**Figure 8:** (a) *Schematic illustration of the Zeeman sublevels in the D1 manifold of $^{87}$Rb atoms that can be used to implement the proposed schemes. (b) Schematic illustration of the $^{87}$Rb transitions that can be used for demonstrating the LSIIB effect in a single cluster without a surrounding cavity. See text for details.*

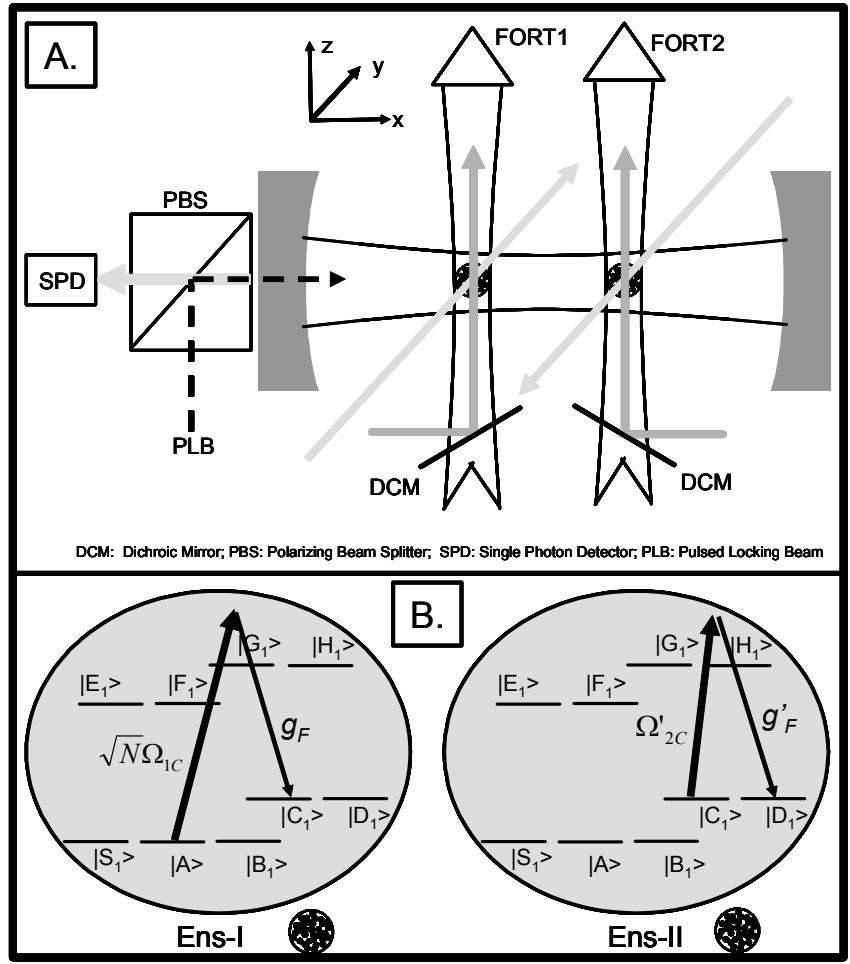

**Figure 9:** *(a) Schematic illustration of the configuration that can be employed for testing the CNOT operation using the ensemble LSIIB. (b) Illustration of the excitations that could be employed for confirming the C-NOT gate operation. See text for details.*

# References:

[1] R.H. Dicke, "Coherence in spontaneous radiation processes", Phys. Rev. 93, 99(1954).

[2] L.M. Duan, M.D. Lukin, J.I. Cirac, and P. Zoller, "Long-distance quantum communication with atomic ensembles and linear optics", Nature 414, 413 (2001).

[3] L.M. Duan, J.I. Cirac, and P. Zoller, "Three-dimensional theory for interaction between atomic ensembles and free-space light", Phys. Rev. A 66, 023818 (2002).

[4] G. K. Brennen, I. H. Deutsch, And P.S. Jessen, "Entangling Dipole-Dipole Interactions For Quantum Logic With Neutral Atoms," Physical Review A, Volume 61, 062309 (2000).

[5] M.D. Lukin, M. Flieschhauer, R. Cote, L.M. Duan, D. Jacksch, J.I, Cirac, and P. Zoller, "Dipole blockade and quantum information processing in mesoscopic atomic ensembles", Phys. Rev. Lett. 87, 037901 (2001).

[6] M.S. Shahriar, P. Pradhan, G.S. Pati, and K. Salit "Light-Shift Imbalance Induced Blockade of Collective Excitations Beyond the Lowest Order**,"** submitted for publication. [ *preprint, available at http://lapt.ece.northwestern.edu/files/LSIIB-for-Collective-Excitation* and at http://arxiv.org/abs/quant-ph/0604120]

[7] T. Pellizzari, S.A. Gardiner, J.I. Cirac, and P. Zoller, " Decoherence, continuous observation, and quantum computing: a cavity QED model", Phys. Rev. Lett. 75 3788 (1995).

[8] S. Lloyd, M.S. Shahriar, J.H. Shapiro, and P.R. Hemmer, "Long Distance, Unconditional Teleportation of Atomic States via Complete Bell State Measurements," Phys. Rev. Lett. **87**, 167903 (2001).

[9] S. Lloyd, J. Shapiro, F. Wong, P. Kumar, M.S. Shahriar, and H. Yuen, "Infrastructure for the Quantum Internet," ACM SIGCOMM Computer Communication Review (Oct. 2004).

[10] D. Gottesman and I. Chuang "Demonstrating the Viability of Universal Quantum Computation using teleportation and single-qubit operations," Nature (London) **402**, 392 (1999).

[11] J.A. Sauer, K.M. Fortier, M.S. Chang, C.D. Hamley, and M.S. Chapman, "Cavity QED with optically transported atoms", Phys. Rev. A 69, 051804 (2004).

[12] W. Alt, D. Schrader, S. Kuhr, M. Muller, V. Gomer, and D. Meschede, "Single atoms in a standing-wave dipole trap", Phys. Rev. A 67, 033403 (2003).

[13] P. Kruger, X. Luo, M.W. Klein, K. Brugger, A. Haase, S. Wildermuth, S. Groth, I. Bar-Joseph, R. Folman, and J. Schmiedmayer, " Trapping and manipulating neutral atoms with electrostatic fields", Phys. Rev. Lett. 91, 233201 (2003).

[14] R. Dumke, M. Volk, T. Muther, F.B.J. Buchkremer, G. Birkl, and W. Ertmer, "Micro-optical realization of arrays of selectively addressable dipole traps: a scalable configuration for quantum computation with atomic qubits", Phys. Rev. Lett. 89, 097903 (2002).

[15] G. Birkl, F.B.J. Buchkremer, R. Dumke, and W. Ertmer, "Atom optics with microfabricated optical elements", Opt. Commun. 191, 67 (2001).

[16] A. V. Turukhin, V.S. Sudarshanam, M.S. Shahriar, J.A. Musser, B.S. Ham, and P.R. Hemmer, "Observation of Ultraslow and Stored Light Pulses in a Solid," *Phys. Rev. Lett.* **88**, 023602 (2002).

[17] M. S. Shahriar, P. R. Hemmer, S. Lloyd, P. S. Bhatia, and A. E. Craig, "Solid-state quantum computing using spectral holes," Physical Review A **66**, 032301 (2002).

[18] M.S. Shahriar, J. Bowers, S. Lloyd, P.R. Hemmer, and P.S. Bhatia, "Cavity Dark State for Quantum Computing,"*Opt. Commun.* **195**, 5-6 (2001)

[19] This can be seen as follows. For a Gaussian cavity mode, the effective mode diameter increases away from the center. As such, the amplitude of the electric field also drops off, in a manner so that the system is equivalent to an uniform intensity over the whole length of the cavity, with an effective diameter that is the same as that at the center.

[20] C.J. Hood, T.W. Lynn, A.C. Doherty, A.S. Parkins, and H.J. Kimble, Science 287, 1447 (2000).

[21] G. Rempe, R. J. Thompson, H. J. Kimble, and R. Lalezari, Opt. Lett. 17, 363(1992).